\documentclass[ALICE,manyauthors]{cernphprep}

\usepackage[comma,square,numbers,sort&compress]{natbib}
\usepackage{hyperref}
\usepackage{lineno}

\usepackage{graphicx,type1cm,eso-pic,color}
\usepackage{epsfig}
\usepackage{graphicx}
\usepackage{fancyhdr}
\usepackage{pslatex}   
\usepackage{color}
\usepackage{amsmath, amsthm, amssymb} 
\usepackage{endnotes}
\usepackage{xspace}
\usepackage{multirow}
\usepackage{epstopdf}
\RequirePackage[normalem]{ulem} 
\RequirePackage{color}\definecolor{RED}{rgb}{1,0,0}\definecolor{BLUE}{rgb}{0,0,1} 
\providecommand{\DIFadd}[1]{{\protect\color{blue}\uwave{#1}}} 
\providecommand{\DIFdel}[1]{{\protect\color{red}\sout{#1}}}                      

\input{definitions.tex}

\begin{document}%

\begin{titlepage}
\PHyear{2016}
\PHnumber{322}      
\PHdate{23 December}  
%

\title{Insight into particle production mechanisms via angular correlations of identified particles in pp collisions at $\bf \sqrt{s}=7$~TeV}

\ShortTitle{Insight into particle production via angular correlations}

\Collaboration{ALICE Collaboration\thanks{See Appendix~\ref{app:collab} for the list of collaboration members}}
\ShortAuthor{ALICE Collaboration} 

\begin{abstract}

Two-particle angular correlations were measured in pp collisions at $\sqrt{s} = 7$~TeV for pions, kaons, protons, and lambdas, for all 
particle/anti-particle combinations in the pair.
Data for mesons exhibit an expected peak dominated by effects associated with mini-jets and are well reproduced by general purpose Monte Carlo generators. However, for baryon--baryon and anti-baryon--anti-baryon pairs, where both particles have the same baryon number, a near-side anti-correlation structure is observed instead of a peak. This effect is interpreted in the context of baryon production mechanisms in the fragmentation process. It currently presents a challenge to Monte Carlo models and its origin remains an open question.


%

\end{abstract}
\end{titlepage}
\setcounter{page}{2}

\section{Introduction}
\label{sec:intro}

Ultrarelativistic proton-proton collisions at the Large Hadron Collider
(LHC) provide a unique opportunity to study Quantum
Chromodynamics (QCD) at new energy scales.
Two-particle angular correlations \cite{Alver:2008aa,Abelev:2009af,Alver:2009id,Khachatryan:2010gv,Aaboud:2016yar,Chatrchyan:2011eka,Aad:2012gla,Aad:2014lta,Chatrchyan:2012wg,CMS:2012qk,Khachatryan:2016txc} are a robust tool  which allows for exploration of the underlying physics phenomena of particle production in
collisions of both protons and heavy ions by measuring the distributions of angles in \detadphi~space (where $\Delta\eta$ is the pseudorapidity difference and $\Delta\varphi$ is the azimuthal angle difference between two particles).
These correlations
open up the possibility to study a number of mechanisms
simultaneously. The baseline, the physics mechanism underlying all correlations, is the global conservation of energy and momentum  as well as strangeness, baryon number, and electric charge. 
It results in a ``$-\cos(\Delta\varphi)$"-like shape spanning the entire phase-space~\cite{Janik:2016nbn}.
Other phenomena, including mini-jets, elliptic flow, Bose-Einstein
correlations, resonance decays,
are sources of additional correlations and each produces a characteristic distribution in \detadphi~space. 
Together with the baseline, they determine the final shape of the correlation function.

This work presents an extension to the traditional angular
correlation measurements, which were studied at the LHC for all avaliable collision systems~\cite{Khachatryan:2010gv,CMS:2012qk,Abelev:2012ola,Aad:2012gla,Adam:2015bka,Adam:2016tsv,Adam:2016ckp}. It is performed for identified particles,
that is pions, kaons, protons, and lambda baryons, produced in proton-proton collisions at $\sqrt{s}=7$~TeV recorded by ALICE~\cite{Adam:2015qaa}. 
The high collision energies provided by the LHC, leading to large cross-sections for production of (anti-)baryons, enable the measurement of correlations not only of primary protons, but also of lambdas and anti-lambdas with very high precision.
Choosing specific particle types allows for the selection of a specific combination of
of quantum numbers (strangeness, baryon number) that may manifest in the measured correlation. As a consequence the
angular correlations for identified particles may reveal new
structures, which reflect the specific conservation laws for these
quantum numbers. The correlations should also be sensitive to details of
particle production mechanisms, including the parton fragmentation. In order to interpret the data in this context, dedicated Monte Carlo simulations using PYTHIA and PHOJET generators were performed.

The paper is organized as follows. Experimental setup, data taking conditions, and track selection criteria are presented in Section~\ref{sec:data}. Section~\ref{sec:analysis} presents the analysis procedure and introduces the correlation function. In Section~\ref{sec:systunc} systematic uncertainties are discussed. The experimental results are presented in Section~\ref{sec:results}. Detailed studies and the comparison of results to Monte Carlo simulations are shown in Section~\ref{sec:mcresults}. Section~\ref{sec:conclusions} summarizes the paper.

\section{Data taking and track reconstruction}
\label{sec:data}
This study was performed on a data sample of about $2.5\times10^8$ pp~events at $\sqrt{s}=7$~TeV recorded by ALICE~\cite{Aamodt:2008zz} in 2010. The trigger system is described in details in Ref.~\cite{Krivda:2012zz}. The minimum-bias trigger required a signal in either V0 (a detector made of the two arrays of scintillating counters V0A and V0C; see Ref.~\cite{Cortese:2004aa} for details) or one of the two inner layers of the Silicon Pixel Detector (SPD), which cover pseudorapidity ranges of $2.8<\eta<5.1$ for V0C and $-3.7<\eta<-1.7$ for V0A and  $|\eta|<1.4$ for the SPD. Each event (collision) consists of global properties and trajectories of particles (tracks).

Two main subsystems were used for particle trajectory reconstruction: the Inner Tracking System (ITS), a silicon-based tracker consisting of 6 layers, and the Time Projection Chamber (TPC). Their acceptance covers the full azimuth within the pseudorapidity range $|\eta|<0.9$. The collision-vertex position was determined with tracks reconstructed in the ITS and the TPC as described in Ref.~\cite{Abelev:2014ffa}. Particle trajectories are reconstructed from a collection of space points. The ITS provides up to 6 points, one at each layer. The TPC provides up to 159 points (clusters), which also contain information on the ionization energy. This information is averaged over all clusters, giving a measurement of the specific ionization energy loss  $\langle \mathrm{d}E/\mathrm{d}x \rangle$ for the particle. For each track, the $\langle \mathrm{d}E/\mathrm{d}x \rangle$ signal is computed as a truncated mean of the lowest 60\% of the all measured points. The charged particle momentum is determined on the basis of the charged particle trajectories bent by the magnetic field of \mbox{0.5 T} parallel to the beam axis.

ALICE provides a particle identification (PID) capability through the combination of the measurement of the specific ionization in the TPC and the timing signals in the Time-Of-Flight (TOF) detector~\cite{Abelev:2014ffa,Abelev:2013vea}. Particle trajectories are propagated from the TPC to the TOF and matched to hits in this detector. Each hit is associated with the time of its detection. The start-time of the event is determined by combining the time estimated using the particle arrival times at the TOF and the time measured by the T0 detector~\cite{Cortese:2004aa,Adam:2016ilk}. This time, combined with the time of the detection of the particle and with the total length of the track enables the measurement of a particle's velocity. This, combined with the momentum obtained from the TPC, enables the determination of particle's mass and therefore its identity.

The protons analyzed here have a minimum transverse momentum of $p_{\mathrm{T}}=0.5$~GeV/$c$, while the kaons and pions are measured for $p_{\mathrm{T}}>0.3$~GeV/$c$ and
$p_{\mathrm{T}}>0.2$~GeV/$c$, respectively.
At lower values of transverse momenta, the sample can be contaminated by particles from the detector material and the PID procedures are also less reliable.
The upper limit of \pt\ for all particle
species was set to 2.5 GeV/$c$; above this value the selected identification
process was no longer able to efficiently distinguish pions from kaons
and protons. 
In order to ensure that the sample of accepted tracks corresponds mostly to primary particles, a $p_{\mathrm{T}}$-dependent selection on the
Distance of Closest Approach (DCA) to the primary vertex was applied. Accepted tracks were located
at a distance  in the transverse plane smaller than $\left(0.018+0.035 p_{\mathrm{T}} ^{-1.01} \right)$ cm, where \pt\ is expressed in GeV/$c$ (which corresponds to about $7\sigma$ of the track DCA resolution), and $2.0$~cm in the beam direction. The contamination from secondary particles was estimated using PYTHIA Perugia-0 \cite{Sjostrand:2006za,Skands:2010ak} simulations and found to be below 1\% for pions and kaons, and below 4\% for protons.
To ensure that only tracks with sufficient reconstruction quality were used in the analysis,
they were required to have a minimum of 70 TPC clusters (maximum possible number is 159)
associated to them; additionally, the maximum value of $\chi^2$ of the Kalman fit per
TPC cluster was set to $4.0$ (2 degrees of freedom per cluster)~\cite{Abelev:2013wsa}.
The selection criteria described above are summarized in Tab.~\ref{tab:V0cuts}.

The particle identification of pions, kaons, and protons was performed on a track by track basis using information from the TPC and TOF detectors; namely the measured values of $\langle \mathrm{d}E/\mathrm{d}x \rangle$, the particle velocity $\beta$, and their resolutions defined as standard deviations around the nominal signal, $\sigma_{\rm TPC}$ and $\sigma_{\rm TOF}$ respectively.
Based on the difference (expressed in units of the resolution $\sigma$) between the measured signal and the expected signal for pions, kaons, or protons in the TPC and TOF, three values of ${N_{\sigma, \rm PID}^{a}}$ (where ${N_{\sigma}}$ is the number of standard deviations of the observable away from the mean 
 and ``a'' denotes one of the three particle type hypotheses) were used to select each track~\cite{Abelev:2013wsa}.
For particles with \pt\ $>$  0.5 GeV/$c$, the values were calculated from the combined TPC--TOF information, ${N_{\sigma, \rm PID}^{a}}^2 = {N_{\sigma, \rm TPC}^{a}}^2 + {N_{\sigma, \rm TOF}^{a}}^2$, resulting in a circular cut in the $N_{\sigma, \rm TPC}^{a}$ and $N_{\sigma, \rm TOF}^{a}$ space.
For \pt\ less than 0.5 GeV/$c$, only a few tracks have an associated signal in the TOF and information only from the TPC was used ($N_{\sigma, \rm PID}^{a} = N_{\sigma, \rm TPC}^{a}$).
In order to form a sample consisting of particles of a species ``a'', particles were selected with $N_{\sigma, \rm PID}^{a} < 2$. Additionally, to keep  the purity of the sample above 96\%, in regions where the areas of two species overlap, an exclusive identification was used. Specifically, tracks  for which the ${N_{\sigma, \rm PID}^{a}} < 3$ condition is fulfilled for more than one particle species hypothesis, were rejected. This procedure resulted in a purity above 99\% for pions and protons and above 96\% for kaons.

The weakly decaying lambda baryons were reconstructed using
their distinctive V-shaped decay topology in the channel 
$\Lambda(\overline{\Lambda})\rightarrow\mathrm{p}\pi^-(\overline{\mathrm{p}}\pi^+)$, which has a branching ratio of 63.9\%~\cite{Olive:2016xmw}.
The reconstruction method forms $V^0$ decay candidates\footnote{$V^0$ candidate is a combination of two secondary tracks of opposite charge which have  a sufficiently large impact parameter with respect to the primary vertex.}; details are described in~\cite{Aamodt:2011zza,Abelev:2013haa}. 
The selection criteria used in this analysis are listed in Tab.~\ref{tab:V0cuts}. They were also varied to estimate the $V^0$ selection systematic uncertainty.
Only $\Lambda$ candidates within an invariant mass window $m_{\Lambda_{\rm PDG}}-0.0038 < m_{V^0} < m_{\Lambda_{\rm PDG}}+0.0038$ GeV/${c}^2$ and $p_{\rm T}>0.6$~GeV/$c$ were used. 
To calculate the $\Lambda$ purity, the signal $S$ was first approximated by
a Gaussian superimposed on a second order polynomial background $B$~\cite{Aamodt:2011zza}.
The $\Lambda$ purity (defined as $S$/$(S+B)$), in the invariant mass range defined above,
was found to be larger than 95\%. 
With respect to the method used in Refs.~\cite{Aamodt:2011zza,Abelev:2013haa} an additional selection on maximum DCA of the $\Lambda$ candidate to the primary vertex was applied to minimize the contribution of weak decays of the charged and neutral $\Xi$. The resulting contamination was estimated using a data-driven approach, 
based on the fit of the $V^0$ transverse DCA to the primary
vertex distribution with the expected shapes for primary
and secondary particles~\cite{Abelev:2013vea}. The contamination corresponds to 17\% of the $V^0$ candidates in the sample.
Another cut was used to prevent two reconstructed $V^0$s from sharing the same daughter track.
If two $V^0$ candidates shared a daughter track, the $V^0$ candidate with the lowest DCA to the primary vertex was chosen, while the other one was rejected from the analysis. 

\begin{table}[t!]
\centering
\begin{tabular}{l|l}
  \hline
 	        Selection variable & Cut value \\ \hline 
 	 \multicolumn{2}{c}{Common track selections} \\ \hline
 	   $|\eta|$ & $\leq 0.8$ \\
 	   Number of TPC clusters & $\geq 70$  \\ 
 	   $\chi^2$ per cluster & $\leq 4$ \\ \hline
     \multicolumn{2}{c}{Primary track selections} \\ \hline
      $\rm DCA_{xy}$ to primary vertex & $\leq \left( 0.0182+0.0350 p_{\mathrm{T}} ^{-1.01} \right)$ cm \\ 
         $\rm DCA_{z}$ to primary vertex & $\leq$ 2 cm \\ 
       Particle identification &  $N_{\sigma, \rm PID} < 2$, with additional rejection, see text  \\ \hline
      \multicolumn{2}{c}{Secondary track selections ($\Lambda$ and $\overline{\Lambda}$ daughters)} \\ \hline

       DCA of daughter track to primary vertex & $\geq 0.06$ cm \\ 
       DCA between daughter tracks & $\leq 1$ cm \\ 
       p ($\rm\overline{p}$) daughter & $0.3\leq p_{\rm T} \leq 4.0$ GeV/$c$ \\ 
       $\pi^{-}$ ($\pi^{+}$) & $0.16 \leq p_{\rm T} \leq 4.0$ GeV/$c$  \\ 
       Particle identification & $N_{\sigma,\rm TPC}< 5$ \\  \hline  
      \multicolumn{2}{c}{$V^0$ vertex selections ($\Lambda$ and $\overline{\Lambda}$)} \\ \hline   
      $|\eta|$ & $\leq 0.8$ \\ 
       DCA of $V^0$ to primary vertex & $\leq 0.6$ cm \\
       Cosine of $V^0$ pointing angle & $\geq 0.99$ \\ 
   $\Lambda$ mass acceptance window & $m_{\Lambda_{\rm PDG}}-0.0038 \leq m_{V^0} \leq m_{\Lambda_{\rm PDG}}+0.0038$ GeV/${c}^2$ \\ 
  $\rm K^0_s$ mass rejection window &  $m_{V^0} \notin $ $({m_{K_{s,  \rm PDG}^0 }}-0.01 ,  {m_{K_{s,  \rm PDG}^0}}+0.01)$ GeV/${c}^2$ \\ 
        \hline
\end{tabular}
\caption{Track and secondary vertex selection criteria.}
\label{tab:V0cuts}
\end{table}

For pairs of pions, kaons and protons, a dedicated procedure (called \emph{share fraction} rejection)
was used to mitigate the effects of merging (two tracks reconstructed as one),
and splitting (one track reconstructed as two). A \emph{share fraction} value for a pair is
obtained as a ratio of the number of times the two tracks share a cluster to the number of all clusters of both tracks in the TPC. All pairs for which this fraction was larger than $5 \%$ 
were rejected.  
In the $\Lambda$ analysis, two main particle pair selection criteria were used. To
resolve two-track inefficiencies associated with daughter tracks,
such as the splitting or merging of tracks discussed above, a
separation cut was employed (the \emph{share fraction} cut is defined for primary tracks only): for each pair,
the spatial separation between the same-sign daughters was
measured at several points throughout the TPC (every 20 cm radially
from 80 cm to 250 cm) and averaged. 
If the average separation of either pair of like-sign tracks was below
3 cm, the $\Lambda$ pair was not used. 
Moreover, the process of photon conversions to $e^{+}e^{-}$ pairs, which produce a very sharp peak at
$(0,0)$, is usually treated as non-interesting in angular
correlation analyses\footnote{By the conservation of momentum and energy, electron positron pairs coming from gamma conversion move parallel to each other (both $\Delta\eta$ and $\Delta\varphi$ are equal to zero). The angular width of the peak is only slightly increased due to the detector resolution.}. In this work their influence was intentionally
minimized with dedicated experimental selection criteria, which remove electron-positron pairs with invariant mass $m_{e^{+}e^{-}}<2$~MeV/$c$ and polar angle difference $\Delta\theta<0.008$ rad.

\section{Analysis}
\label{sec:analysis}
The correlation function for two particles,  of type ``1" and type ``2" (which in general may be non-identical), is expressed as the inclusive two-particle distribution $P_{12}(\varphi_1,\eta_1,\varphi_2,\eta_2)$ divided by the product of the inclusive single-particle distributions $P_1(\varphi_1,\eta_1)$ and $P_2(\varphi_2,\eta_2)$\footnote{We note that the expression given by Eq.~(\ref{eq:corrfundef}) is general and also applicable to all two-particle correlations by substituting $\varphi$ and $\eta$ with the quantities of interest.}:
\begin{equation}
C(\varphi_1,\eta_1,\varphi_2,\eta_2)=\frac{P_{12}(\varphi_1,\eta_1,\varphi_2,\eta_2)}{P_1(\varphi_1,\eta_1)P_2(\varphi_2,\eta_2)}.
\label{eq:corrfundef}
\end{equation}
We can interpret the distribution $P_{12}(\varphi_1,\eta_1,\varphi_2,\eta_2)$ as a conditional probability to observe a particle with azimuthal angle $\varphi_1$ and pseudorapidity $\eta_1$ if a particle with azimuthal angle $\varphi_2$ and pseudorapidity $\eta_2 $ is observed as well, and distributions $P_1(\varphi_1,\eta_1)$ and $P_2(\varphi_2,\eta_2)$ as probabilities of observing particles with $\varphi_1$ and $\eta_1$, and $\varphi_2$ and $\eta_2$, respectively. In the absence of correlations, the ratio should equal unity.

Therefor,e the  experimental correlation function, including physical two-particle
correlations as well as effects from single-particle acceptance, is constructed as

\begin{equation}
\label{eq:CorrelationFuntion}
C(\Delta\eta,\Delta\varphi)=\frac{S(\Delta\eta,\Delta\varphi)}{B(\Delta\eta,\Delta\varphi)},
\end{equation}
where $\Delta\eta=\eta_1 - \eta_2$ is the difference in
pseudorapidity, $\Delta\varphi=\varphi_1 - \varphi_2$ is the
difference in azimuthal angle, $S(\Delta\eta,\Delta\varphi)$ is the distribution of correlated pairs and $B(\Delta\eta,\Delta\varphi)$ is the reference distribution, reflecting the single-particle
acceptance. $S$ is constructed from particle pairs
coming from the same event
\begin{equation}
\label{eq:Nsignal}
S(\Delta\eta,\Delta\varphi)=\frac{1}{N_{pairs}^{signal}}\frac{\mathrm{d}^2N_{pairs}^{signal}}{\mathrm{d}\Delta\eta \mathrm{d}\Delta\varphi},
\end{equation}
where $N_{pairs}^{signal}$ is the number of pairs of particles in $S$. $B$ is
constructed using an event mixing technique, where each particle
in the pair comes from a different event and can be expressed as
\begin{equation}
\label{eq:Nmixed}
B(\Delta\eta,\Delta\varphi)=\frac{1}{N_{pairs}^{mixed}}\frac{\mathrm{d}^2N_{pairs}^{mixed}}{\mathrm{d}\Delta\eta \mathrm{d}\Delta\varphi},
\end{equation}
where $N_{pairs}^{mixed}$ is the number of pairs of particles in
$B$.
Given each of the $S$ and $B$ distributions are divided by the respective number of pairs, the reported distribution is a ratio of probabilities, as defined in Eq. (\ref{eq:corrfundef}).  

In order to improve the reference estimation, particles from
each event, selected according to Tab.~\ref{tab:V0cuts}, are combined with particles from other events, for which the multiplicities differ by no more than 5 tracks and primary vertex positions differ by no more than 2 cm in the beam direction. Each event is mixed with 10 events.

Each particle is weighted with a correction factor that accounts for
detector acceptance, reconstruction and particle identification efficiencies, as well as  contamination by secondary particles calculated using events from PYTHIA6 (Perugia-2011 tune) with particle transport performed via a GEANT3 \cite{Brun:1994aa} simulation of the ALICE detector.
Applied corrections are \pT-dependent and the correction method is validated on simulated events.\footnote{Correlation functions obtained from MC models after full ALICE reconstruction chain, with all corrections applied, were compared to the correlation functions obtained on the particle level. The correction procedure was tested on PYTHIA6 Perugia-0 and Perugia-2011 tunes.} 
Therefore, the distributions are considered unfolded to the particle level.
The contamination by secondary particles from weak decays (mostly charged and neutral $\Xi$) is estimated by varying the DCA cut and found to be relevant, after applying corrections, only for lambda particles. 
The related systematic uncertainty was calculated from the lambda sample in which the weak decay content was increased by 50\%.
The influence of misidentified particles was estimated by applying alternative PID methods for pions, kaons, and protons (i.e. by increasing the misidentification rate by 100\%) and varying the invariant mass window for lambdas as well as by taking  the misidentification fraction extracted from simulations into account.

\section{Systematic uncertainty}
\label{sec:systunc}
In order to asses the systematic uncertainty connected to the measurement, the selection criteria discussed in Sec.~\ref{sec:data} were modified and for each new set of them the analysis was repeated. The contribution from the different systematic sources were added in quadrature. The correlation functions were found to be rather insensitive to changes of selection criteria, yielding total uncertainty from about 1.5\% for pions to 7.5\% for lambdas.
Due to  different methods of particle reconstruction for neutral lambdas ($V^0$s) and charged pions, kaons, and protons (tracks) the calculation of systematic uncertainty for each of them was done separately.

In the case of evaluation of the tracking uncertainty for pions, kaons, and protons, an alternative track selection was used, where two classes of tracks are combined in order to avoid an azimuthally-dependent tracking efficiency due to inactive SPD modules~\cite{Abelev:2012ej}. The first class requires for tracks to have at least one hit in the SPD (as in the default track selection). For tracks which do not fulfill this criterion, in the second class, the primary vertex position is used as an additional constraint in the global track fit.
To assess the systematic uncertainties related to the particle identification, the misidentification rate was doubled with respect to the values reported in Sec.~\ref{sec:data}. These uncertainties were found to be lower than 0.5\% for pions, 1\% for kaons, and 2\% for protons.
Finally, for pions, kaons, and protons separate analyses were performed for four the datasets collected, changing the single-track inefficiencies. 
For like-sign pairs the results were obtained separately for positive and negative particles.

The systematic uncertainty on $V^0$ selection was evaluated by varying selection criteria discussed in Sec.~\ref{sec:data}; minimum DCA of $V^0$ to the primary vertex, maximal DCA of daughters to the primary vertex, $V^0$ decay length and cosine of pointing angle. Based on Pythia Perugia-2011 calculations, this increased or decreased the default content of fake $V^0$s in the sample of 0.7\% by a factor of 2 or 3, respectively.
The invariant mass window was varied in the systematic uncertainty estimation procedure to $\pm 0.0044$ GeV/${c}^2$. 
To estimate the systematic uncertainty of feed-down from weak decays the selection on the DCA was removed, resulting in an increase in the contamination coming from secondary particles from 17\% to 26\%. The uncertainty of the correction procedure on PID efficiencies and contamination from secondary particles on both tracks and $V^0$s are discussed in Sec.~\ref{sec:analysis}.

Table~\ref{tab:syst} summarizes the main sources of systematic uncertainties of the measurement.

\begin{table}[bht!] \centering
  \begin{tabular}{lcccc}
    \hline
    Source					& pions & kaons & protons & lambdas\\
    \hline
      Track selection and efficiencies 		& $<$0.5\% 	&  $<$2\% 	&  $<$2\% &  -- \\
      Particle identification	 			& $<$0.5\% 	& $<$1\%	& $<$2\%      &  --\\
      Dataset comparison					& $<$1\%	&$<$1\%		& $<$1\%      & -- \\
      $V^0$ selection							& --     	& --    	& --          & $<$5\% \\     	
      $V^0$ signal extraction					& --     	& --    	& --          & $<$4\% \\   
      Feed-down	(weak decays)				& --     	& --    	& --          & $<$4\% \\   
    \hline
      Sum			 						& 1.5\% 	& 	2.5\%	& 3\%       & 7.5 \%\\
    \hline
  \end{tabular}
  \caption{Summary of the main systematic uncertainties.
  \label{tab:syst}
    }
\end{table}

\section{Results}
\label{sec:results}
\begin{figure}[ht!]
	\centering
	\includegraphics[width=0.8\textwidth]{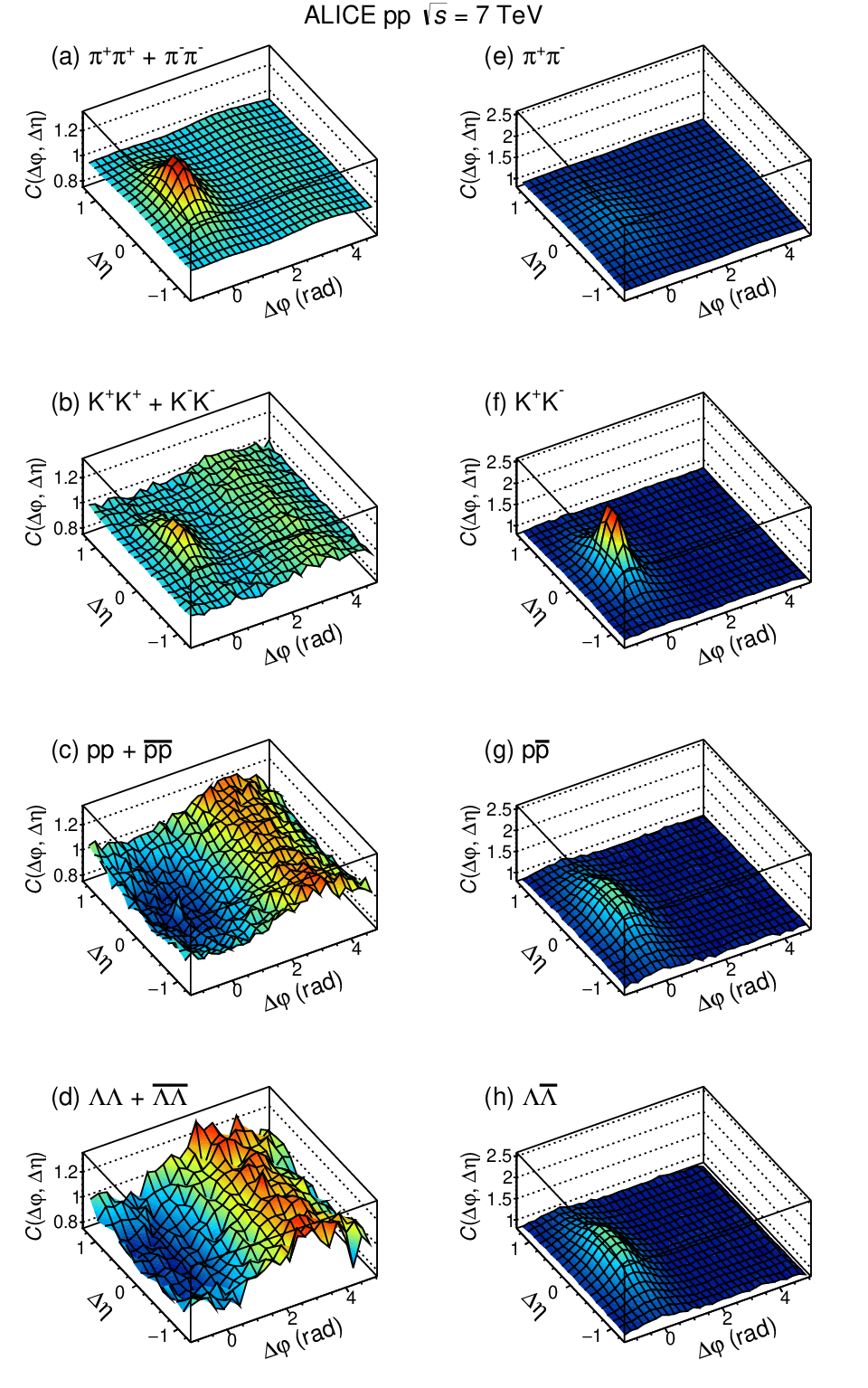}
	\caption{Correlation functions for identical-particle pairs: $\pi^{+}\pi^{+}+\pi^{-}\pi^{-}$, $\rm K^{+}K^{+}+K^{-}K^{-}$, $\rm pp+\overline{p}\overline{p}$, $\Lambda\Lambda+\overline{\Lambda}\overline{\Lambda}$ (left panel) and particle--anti-particle pairs: $\pi^{+}\pi^{-}$, $\rm K^{+}K^{-}$, $\rm p\overline{p}$, $\Lambda\overline{\Lambda}$ (right panel). Plots are mirrored around $\Delta\eta=0$.}
	\label{Fig:corrResultsMixed}
\end{figure}

The measured correlation functions for the four analyzed particle
species (pions, kaons, protons, lambdas) are shown in
Fig.~\ref{Fig:corrResultsMixed}. In the measurement of correlations in pp collisions at LHC energies a distinct near-side peak at $(\Delta\eta,\Delta\varphi)$ about $(0,0)$ is observed  \cite{Khachatryan:2010gv,Khachatryan:2015lva,Aad:2015gqa}, 
which is a combination of at least three effects: (i) fragmentation of hard-scattered partons, (ii) resonance decays, and (iii) femtoscopic correlations.
\mbox{(i) The} fragmentation  originating from low momentum-transfer scatterings, sometimes referred to as mini-jets~\cite{Abelev:2014mva},
produces a broad structure extending at least over one unit in $\Delta\eta$
and $\Delta\varphi$.
(ii) The decay of
resonances contributes to the near-side peak of the correlation function or produces a ridge at $\Delta\eta=0$ (extended in
$\Delta\varphi$~\cite{Eggert:1974ek,Khachatryan:2010gv,Alver:2007wy}), depending on the released kinetic energy of a given resonance. This effect plays a significant role only for correlation functions of unlike-sign particle pairs. 
(iii) The third effect, femtoscopic correlations (an enhancement due to Bose-Einstein quantum statistics for identical bosons, a suppression due to Fermi-Dirac quantum statistics for identical fermions, as well as Coulomb and strong final-state interactions), is present for particles at low relative momenta. The shape of this effect in \detadphi\
depends strongly on the mass of the particle type considered, as well
as on the size of the particle-emitting system. For pp collisions at
ALICE this size was measured in great detail with
pions~\cite{Aamodt:2011kd} and kaons~\cite{Abelev:2012sq}. The
expected width of the correlation peak produced by like-sign charged particles, e.g., pions, is comparable to the
one for the mini-jet peak.
In addition, by
constraints on the energy-momentum conservation, an ``away-side ridge'' structure at $\Delta\varphi=\pi$, with a magnitude only weakly changing with $\Delta\eta$, is produced as well.

For correlation functions (a)-(h) in Fig.~\ref{Fig:corrResultsMixed} the baseline reflecting the energy-momentum conservation is combined with several expected physics mechanisms. For same sign particle pairs of mesons (a)-(b), the near-side peak is consistent with the mini-jet mechanism combined with the
Bose-Einstein correlations. The away-side ridge is also prominent,
consistent with a mini-jet origin.

The particle--anti-particle correlations (e)-(h) also show a mini-jet like
structure on the near-side and a weak away-side one. For pairs of
non-identical particles Bose-Einstein and Fermi-Dirac effects are
not present; however, resonances play a significant role in shaping the correlation function.  
Baryon and meson correlations are qualitatively
similar. The only difference is the magnitude and width of the near-side peak,
which is highest for kaons, lower for protons and lambdas, and lowest
for pions. 
The shape and strength of the correlation functions (a distinct near-side peak) in (a)-(b) and (e)-(h) suggest that they might be dominated by significant mini-jet contributions.

In contrast to like-sign meson correlations, the baryon--baryon (combined with anti-baryon--anti-baryon) distributions for identical proton (c) and lambda (d) pairs show a qualitatively different effect of a wide near-side depression instead of the peak, combined with an away-side
ridge. Such a structure resembles the one associated with the baseline
global energy-momentum conservation. 
Thus, this strong near-side suppression means that the mechanisms which would produce a peak are either not present or produce a very different correlation shape.
On the other hand, a  clear correlation of particles with opposite
baryon number (g)-(h) is observed, resembling the structures observed for unlike-sign mesons.
Based on the results of these studies we can draw the following conclusion: 
if we consider a process of mini-jet fragmentation as the one producing a strong, positive near-side correlation then 
 baryon--anti-baryon pairs  are produced in mini-jets (see (g)-(h)). However, producing more than one such a pair in a single fragmentation is
strongly suppressed (see (c)-(d)).

\begin{figure}[h]
	\centering
	\includegraphics[width=0.8\textwidth]{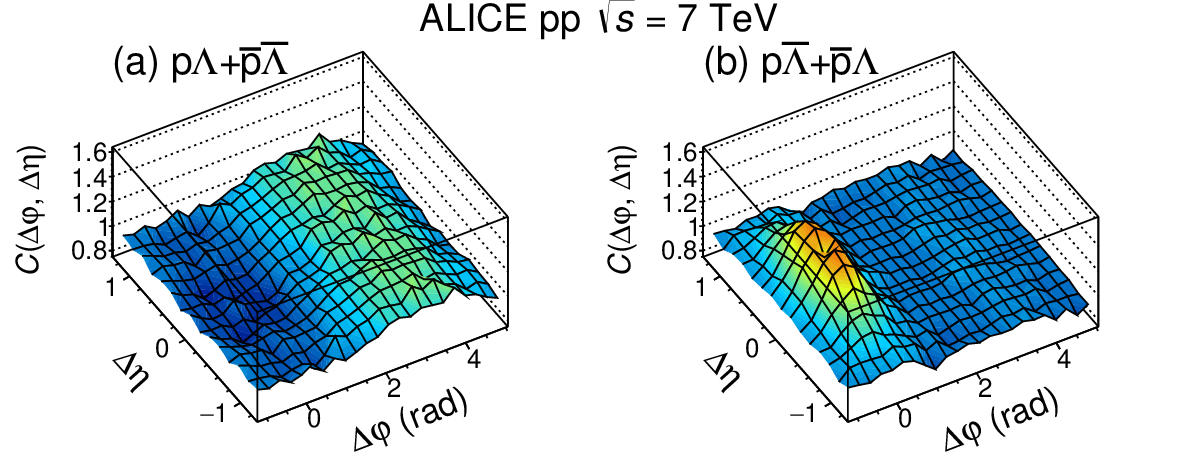}
	\caption{Correlation functions for combined pairs of  (left) $\rm p\Lambda+\overline{p}\overline{\Lambda}$ and (right) $\rm p\overline{\Lambda}+\overline{p}\Lambda$. Plots are mirrored around $\Delta\eta=0$.}
	\label{Fig:PL_APAL_PAL_APL_LHC10bcde}
\end{figure}

There are several hypotheses that could explain the depression observed for baryon--baryon pairs. First, it may be that we are seeing the effects of Fermi-Dirac statistics originating from the wave-function \mbox{(anti-)symmetrization} for identical baryon pairs for triplet (singlet) pair spin combinations~\cite{Aamodt:2011kd,Adamczyk:2015hza}. 
However, using an effective source size of about 1.5 fm (comparable to the measured source size for pions~\cite{Aamodt:2011kd}), the repulsive effects of
Fermi-Dirac statistics should be limited to baryon pairs with momentum differences of less than 200 MeV/$c$ --
a far too short range to explain our observations. The Coulomb and strong final-state interactions would also be convoluted with the quantum statistics effect (the combination of those three effects is referred to as a ``femtoscopic'' effect~\cite{Aamodt:2011kd,Adamczyk:2015hza}). In panel (g) an additional peak at $(0,0)$ with a height of $\sim 0.2$ and a width comparable to the size of a single bin is observed. The origin of that structure was studied by measuring the ``femtoscopic'' correlation function in momentum space and transforming it with a simple Monte Carlo procedure to $(\Delta\eta,\Delta\varphi)$ space. Such procedure shows that this peak can be qualitatively and quantitatively reproduced by ``femtoscopic'' correlations mentioned above. Moreover, the Fermi-Dirac suppression cannot be present for non-identical baryon pairs, like $\rm p\Lambda+\overline{p}\overline{\Lambda}$, which were also measured. The results are shown in Fig.~\ref{Fig:PL_APAL_PAL_APL_LHC10bcde}. One observes that the characteristic shape of anti-correlation is preserved also in this case. 

\begin{figure}[ht]
	\centering
	\includegraphics[width=0.49\textwidth]{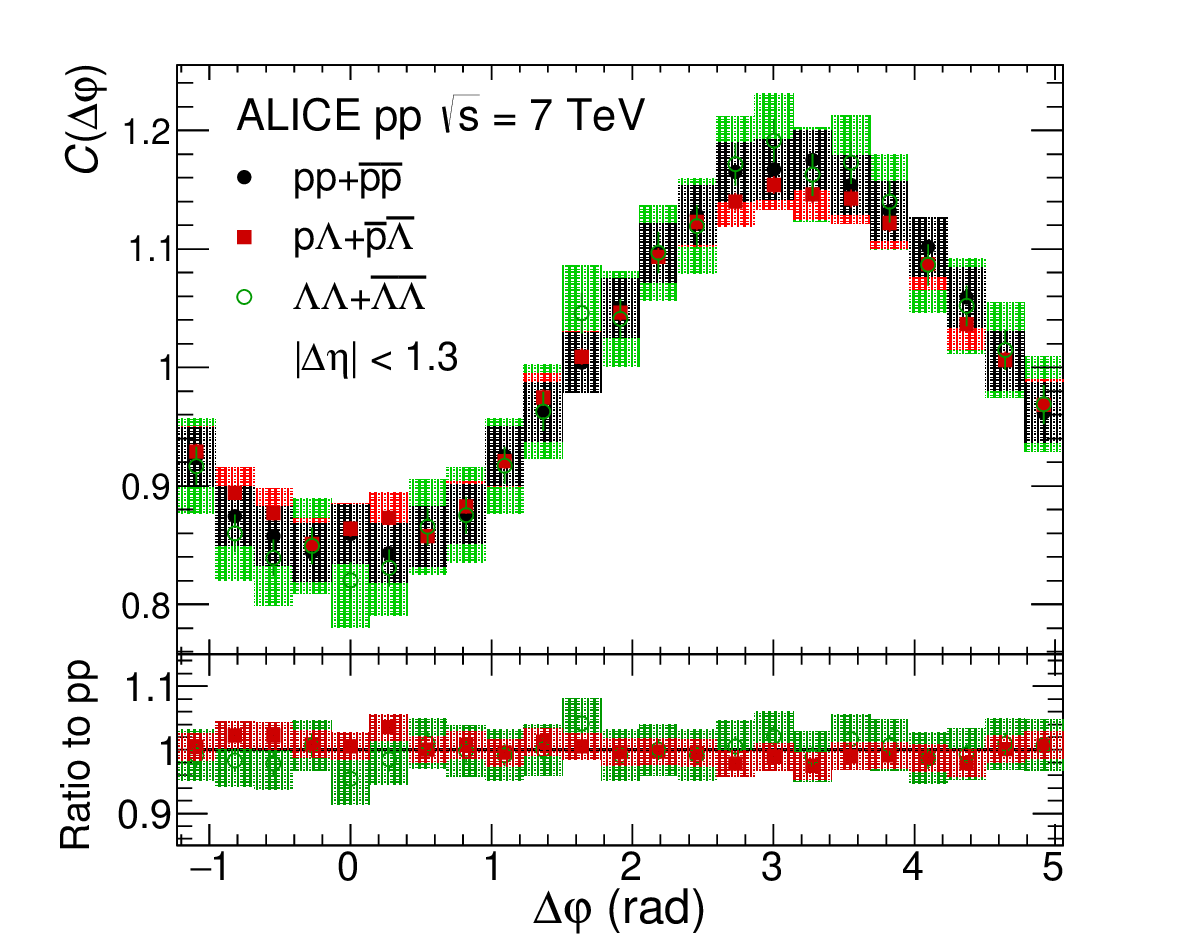}
	\includegraphics[width=0.49\textwidth]{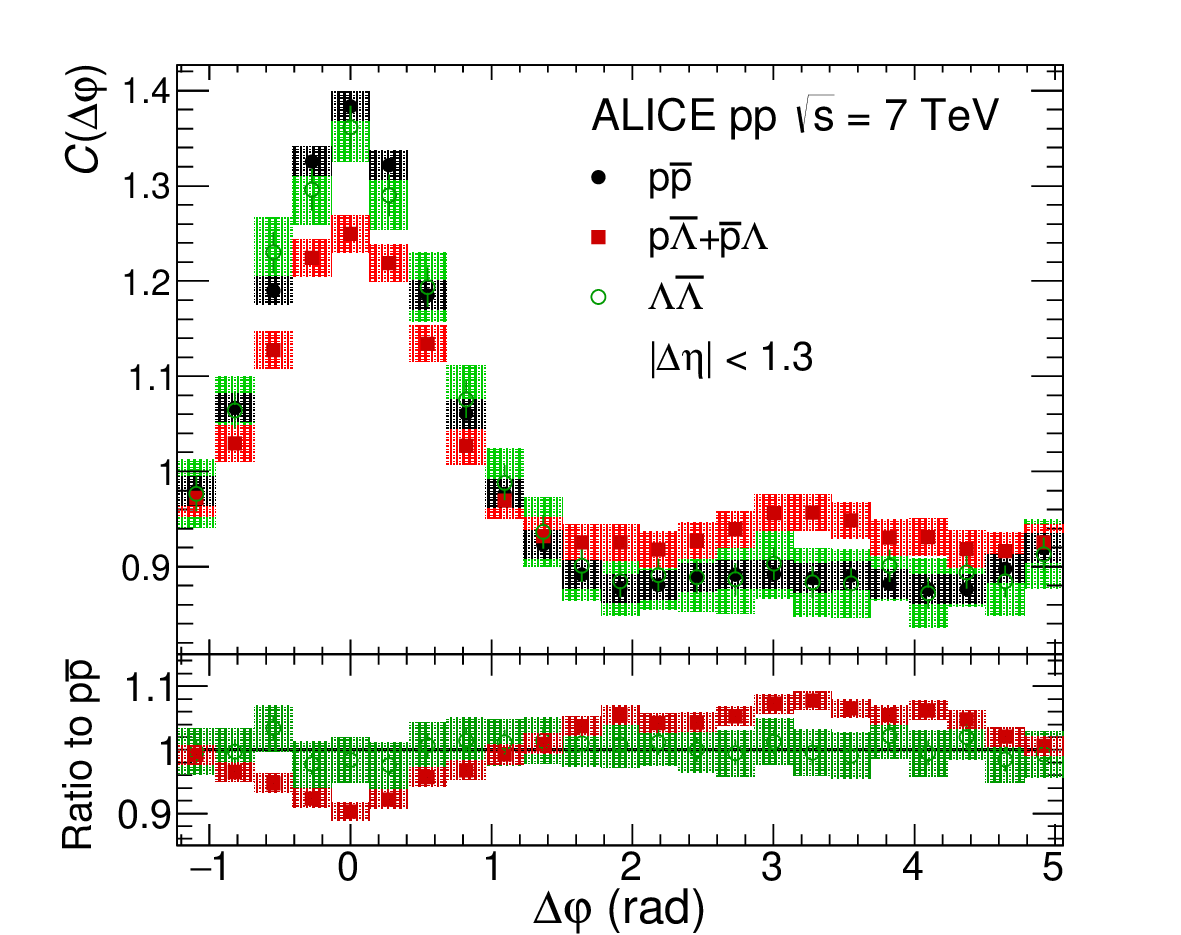}
	\caption{$\Delta\eta$ integrated projections of correlation functions for        combined pairs of (left)  $\rm pp+\overline{p}\overline{p}$, $\rm p\Lambda+\overline{p}\overline{\Lambda}$, and $\Lambda\Lambda+\overline{\Lambda}\overline{\Lambda}$  and (right) $\rm p\overline{p}$, $\rm p\overline{\Lambda}+\overline{p}\Lambda$, and $\Lambda\overline{\Lambda}$. Statistical (bars) and systematic (boxes) uncertainties are plotted.
	}
	\label{Fig:Proj_PAP_PAL_LAL_LHC10bcde}
\end{figure}

The comparison of all baryon pairs, shown as a function of $\Delta\varphi$ and integrated over $\Delta\eta$, can be seen in Fig.~\ref{Fig:Proj_PAP_PAL_LAL_LHC10bcde}. 
The shape of the correlation function for all studied baryon--baryon (and baryon--anti-baryon) pairs is  similar, regardless of particles' electric charge. Therefore, we must reject the hypothesis that Fermi-Dirac quantum statistics is the cause of the observed depression for baryon--baryon pairs. Similar conclusions were reached based on observations of baryon production in \ee\ collisions, see Ref.~\cite{Aihara:1986fy}. The depression is a characteristic attribute connected solely to the baryonic nature of a particle.

\begin{figure}[ht]
	\centering
	\includegraphics[width=.99\textwidth]{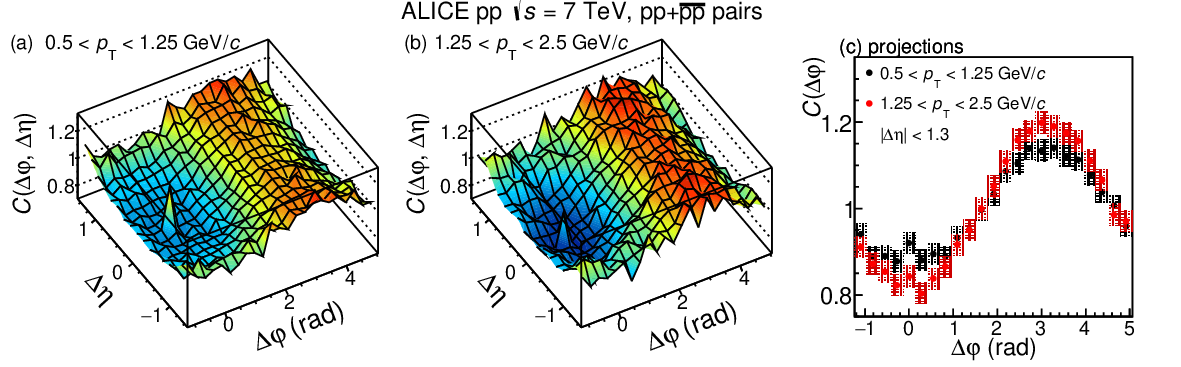}
	\caption{Correlation functions for combined pairs of $\rm pp+\overline{p}\overline{p}$ for two  transverse momentum intervals \mbox{(a) $0.5 <p_{\rm T} < 1.25$ GeV/$c$} and (b) $1.25 <p_{\rm T} < 2.5$~GeV/$c$. Panel (c) shows $\Delta\eta$ integrated projections. Statistical (bars) and systematic (boxes) uncertainties are plotted.}
	\label{Fig:ProtonsPtSum}
\end{figure}

In order to check whether some fraction of the observed effect depends on the momentum transfer during the interaction, the $\rm pp+\overline{p}\overline{p}$ sample was divided into two transverse momentum ranges. 
The correlation functions obtained with these selection criteria are shown in Fig.~\ref{Fig:ProtonsPtSum} and show even stronger anti-correlation for higher transverse momenta of particles in the pair.

An alternative interpretation of the observed depletion is that this structure is the manifestation of a local conservation of baryon number influencing the hadronization process, as argued in Ref.~\cite{Aihara:1986fy} in
the analysis of \ee\ collisions at $\sqrt{s}=29$~GeV. By ``local'' we denote the
production of particles close together in the phase-space
(e.g. in the same mini-jet), as opposed to
``global'' conservation which applies to all particles produced in an
event. In string hadronization models, the ``local'' mechanism requires that two baryons produced in a single fragmentation are separated by
at least one particle with a different baryon number \cite{Aihara:1986fy}.
The production of two baryons in a mini-jet would also be suppressed if
the parton energy is small when compared with the minimum energy
required to produce four baryons (2 particles + 2 anti-particles, the minimum amount to satisfy the law of local baryon number conservation when two baryons are produced in single mini-jet).
At a collision energy of $\sqrt{s}=29$~GeV it was reasonable to assume that the
energy constraint would dominate.
However, at LHC energies this constraint should have less of an impact on the observed structures.
We tested this expectation by employing Monte Carlo generators which
include local baryon number conservation: PYTHIA (6.4 and 8) and
PHOJET (1.12). The results of MC simulations are discussed in the next section.

\section{Comparison to Monte Carlo models}
\label{sec:mcresults}

The correlation functions measured in this work are compared to predictions of Monte Carlo (MC) models. The following MC event
generators were used: PYTHIA6.4 tunes Perugia-0 and Perugia-2011~\cite{Sjostrand:2006za,Skands:2010ak}, PYTHIA8 Monash
tune~\cite{Sjostrand:2007gs,Skands:2014pea} and PHOJET version 1.12~\cite{Engel:1995sb}.
PYTHIA, widely used for simulations of high-energy collisions, combines perturbative QCD for large momentum-transfer interactions and phenomenologically motivated models for the description of soft hadronic interactions; the Lund string fragmentation model \cite{Andersson:1983ia} is used for hadronization. PYTHIA has many free parameters which are optimized to best describe specific measurements. These parameters are collected in predefined ``tunes".  Perugia-0 was tuned for the best description of data up to $\sqrt{s}=1.96$~TeV \cite{Skands:2010ak}. Perugia-2011 takes into account some of the early LHC results at $\sqrt{s}=900$~GeV and 7~TeV, along with increased baryon production (especially of strange baryons) as well as removing suppression of strangeness in fragmentation models~\cite{Skands:2010ak}.
The PYTHIA8 Monash tune~\cite{Sjostrand:2007gs,Skands:2014pea} includes further improvements to the parameters by comparing them with both $e^{+}e^{-}$ collisions and recent pp LHC data, including strange particle and baryon production rates.
The PHOJET generator successfully describes  experimental data measured at collision energies up to $\sqrt{s}=1.8 $~TeV; however, it has not been updated to reproduce LHC data. PHOJET uses the Dual Parton Model \cite{Capella:1992yb} for the simulation of  particle production in low-\pT\ processes and thus it is interesting to consider in addition to PYTHIA.
Similarly to PYTHIA, it incorporates the Lund string fragmentation model.

\begin{figure}[h]
	\centering
	\includegraphics[width=0.88\textwidth]{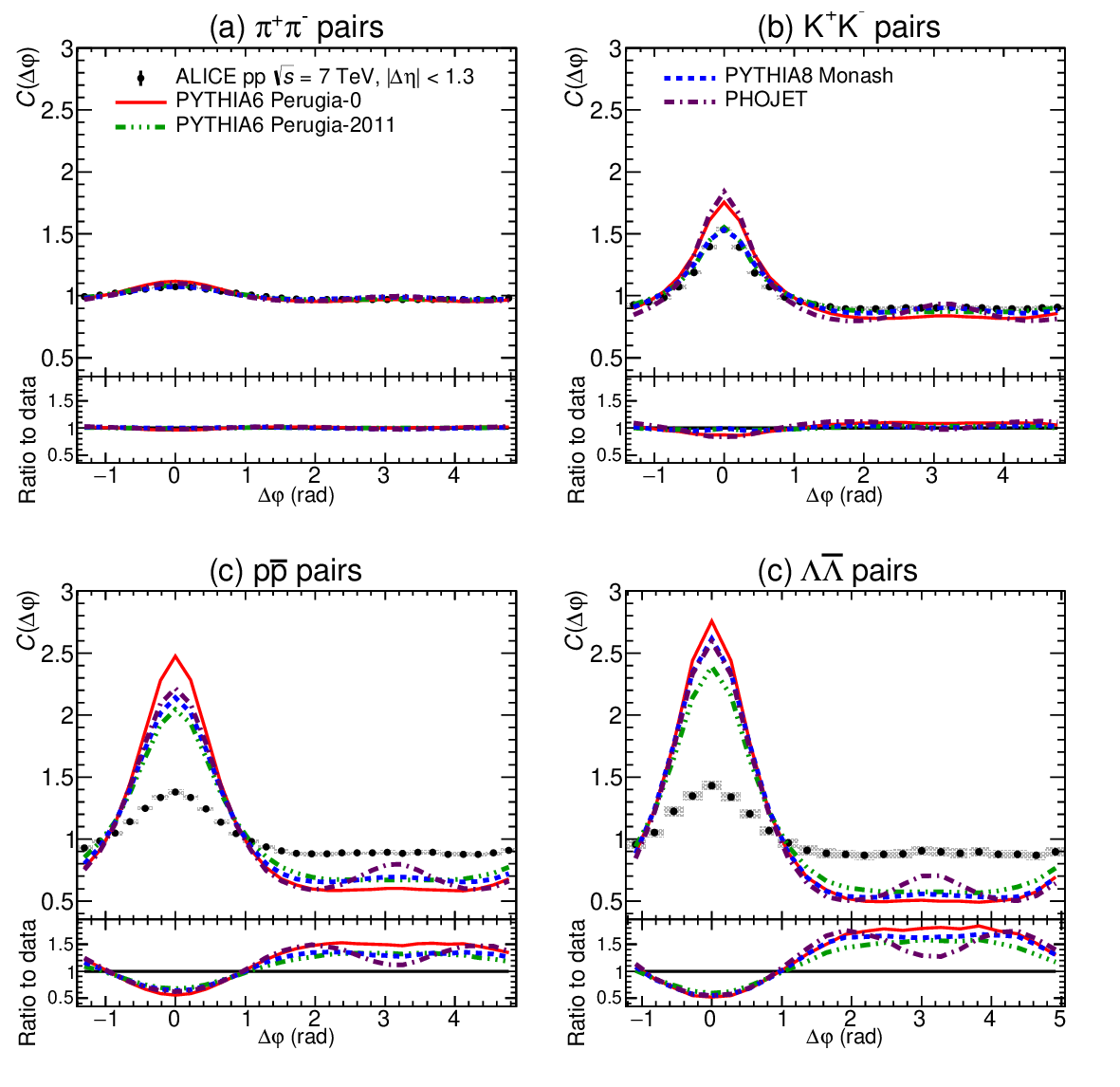}
	\caption{$\Delta\eta$ integrated projections of correlation functions for (a) $\pi^+\pi^-$, (b) $\rm K^+K^-$, (c) $\mathrm{p\overline{p}}$, and (d) $\Lambda\overline{\Lambda}$ pairs obtained from ALICE data and four Monte Carlo models (PYTHIA6 Perugia-0, PYTHIA6 Perugia-2011, PYTHIA8 Monash, PHOJET) at $\sqrt{s}=7\ \rm{TeV}$. Bottom panels show ratios of MC models to ALICE data.  Statistical (bars) and systematic (boxes) uncertainties are plotted.}
	\label{Fig:Proj_CompMC_PIpPIm_KpKm_PAP}
\end{figure}

\begin{figure}[h]
	\centering
	\includegraphics[width=0.88\textwidth]{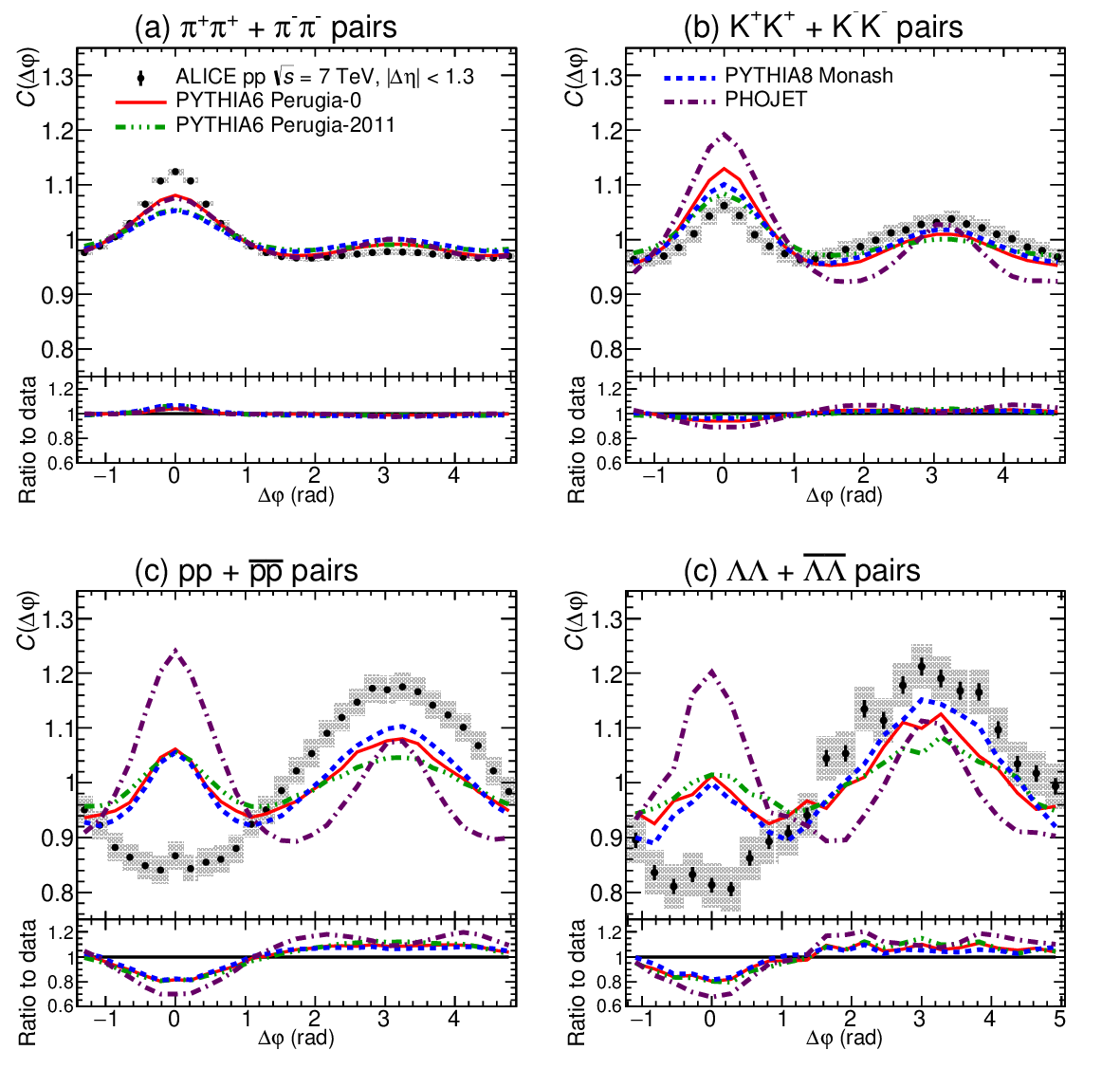}
	\caption{$\Delta\eta$ integrated projections of correlation functions for combined pairs of (a) $\pi^+\pi^+ + \pi^-\pi^-$, (b) $\rm K^+K^+ + K^-K^-$, (c) $\mathrm{pp} + \rm \overline{p}\overline{p}$ and (d) $\Lambda\Lambda+\overline{\Lambda}\overline{\Lambda}$, obtained from ALICE data and four Monte Carlo models (PYTHIA6 Perugia-0, PYTHIA6 Perugia-2011, PYTHIA8 Monash, PHOJET). Bottom panels show ratios of MC models to ALICE data. Statistical (bars) and systematic (boxes) uncertainties are plotted.}
	\label{Fig:Proj_CompMC_PIpPIp_KpKp_PP}
\end{figure}

In Fig.~\ref{Fig:Proj_CompMC_PIpPIm_KpKm_PAP} and Fig.~\ref{Fig:Proj_CompMC_PIpPIp_KpKp_PP} $\Delta\eta$ integrated correlation function projections onto $\Delta\varphi$, integrated over $|\Delta\eta|<1.3$ for particle--anti-particle pairs and for particle--particle pairs (combined with anti-particle--anti-particle pairs) from four different MC calculations are compared to ALICE data.

The MC models reproduce the experimental results reasonably well for mesons. It should be noted that none of the models include quantum-statistics effects, therefore a smaller correlation strength in the near-side region is expected for correlation functions of identical particles in comparison to the experimental data. 
Both tunes of PYTHIA6.4, Perugia-0 and Perugia-2011, give the results which are close to the experimental data (Perugia-0 for pions and Perugia-2011 for kaons). They were also successfully used to describe the non-femtoscopic correlations underlying the Bose-Einstein statistics signal in femtoscopic measurements of identical pions~\cite{Aamodt:2011kd} and identical kaons~\cite{Abelev:2012sq}, respectively. 

However, the models fail to reproduce baryon correlations (both particle--particle and particle--anti-particle pairs). 
No depression is observed for protons and lambdas for any of the studied models. Instead, a near-side peak is present for particle--particle pairs. Furthermore, additional studies were performed, concluding that the anti-correlation cannot be reproduced by tuning parameters of PYTHIA6.4. Apparently all models frequently produce two baryons close in
phase-space (within the mini-jet peak). These results argue against the hypothesis that the combination of energy and baryon-number conservation is enough to explain the observed near-side anti-correlation, since both local baryon number and energy conservation laws are implemented in all studied models.

\begin{figure}[h]
	\centering
	\includegraphics[width=0.88\textwidth]{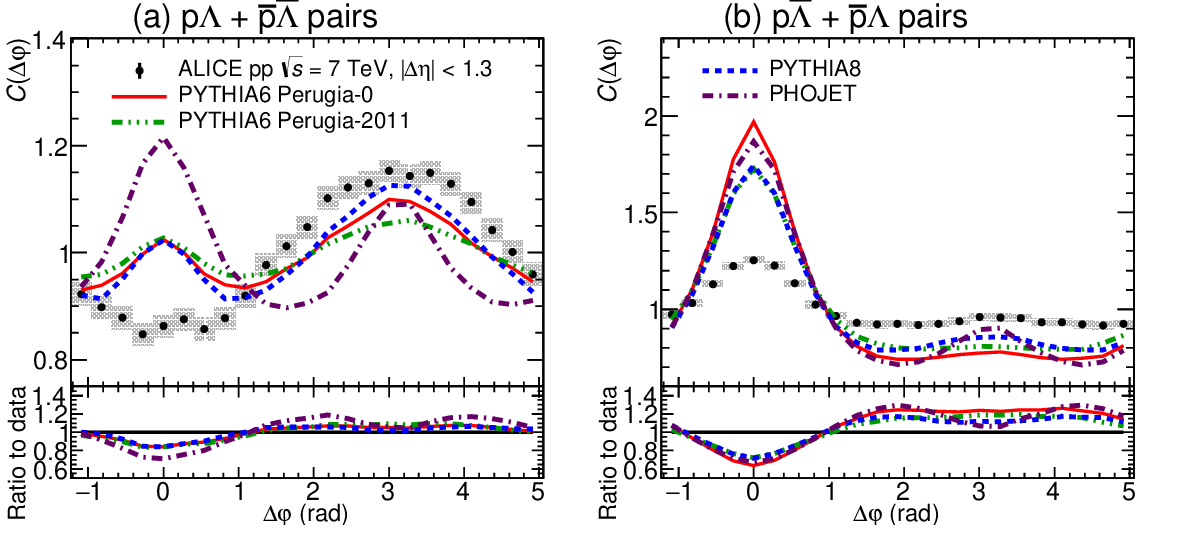}
	\caption{$\Delta\eta$ integrated projection of correlation functions for combined pairs of (left) $\rm p{\Lambda}+\overline{p}\overline{\Lambda}$ and (right) $\rm p\overline{\Lambda}+\overline{p}\Lambda$ obtained from ALICE pp collision data and four Monte Carlo models (PYTHIA6 Perugia-0, PYTHIA6 Perugia-2011, PYTHIA8 Monash, PHOJET) at $\sqrt{s}=7\ \rm{TeV}$. Clear anti-correlation is observed for all baryon pairs. 
	Statistical (bars) and systematic (boxes) uncertainties are plotted.}
	\label{Fig:Proj_CompMC_PAP_PAL}
\end{figure}

For baryons, pronounced differences are also seen for particle--anti-particle pairs; the magnitude of the near-side peak is much higher in all MC models than in ALICE data. 
The universality of this behaviour for all baryon pairs is further confirmed with the studies of the proton--lambda correlations, as shown in Fig.~\ref{Fig:Proj_CompMC_PAP_PAL}. The results show that $\rm p\Lambda+\overline{p}\overline{\Lambda}$ correlation functions follow the trend common for all baryon--baryon pairs, and correlation functions of $\rm p\overline{\Lambda}+\overline{p}\Lambda$ behave similarly to the baryon--anti-baryon correlations.
The away-side correlation is similar to the experimental data for all pair combinations.

\section{Conclusions}
\label{sec:conclusions}

Angular correlations of identified particles were analyzed in \pp~collisions at $\sqrt{s}=7$~TeV recorded with the ALICE experiment. The studies were done separately for particle/anti-particle pairs (for like-sign and unlike-sign pairs) and for four particle species (pions, kaons, protons, lambdas). A significant depression around $(\Delta\eta,\Delta\varphi)\approx(0,0)$ is observed for the baryon--baryon and anti-baryon--anti-baryon pairs, which is not seen for baryon--anti-baryon pairs. 

The analysis was complemented by Monte Carlo model calculations using the PYTHIA6.4 Perugia-0, Perugia-2011, PYTHIA8 and PHOJET (v.\ 1.12), two event generators designed to simulate high momentum fragmentation (i.e. jets). 
While the correlation functions of mesons are well-reproduced by the studied models, those of baryons in simulations are significantly different than those in collision data. The most surprising result is obtained for baryon--baryon (antibaryon--antibaryon) pairs where the models are unable to reproduce even qualitatively the depletion which is observed experimentally. In the case of baryon--antibaryon pairs the correlations are qualitatively comparable, however the simulated ones are much stronger than those observed in collision data.

The observed differences can therefore mean that either the models describe the hadronization process properly but the jet fragmentation is not the dominant mechanism involved in the production of particles found in the measured \pt\ range (\pT\ $<$ 2.5 GeV/$c$), or the fragmentation mechanisms used in PYTHIA and PHOJET are incomplete. The latter scenario would further suggest that some additional, not yet identified mechanism must exist, which suppresses the production of more than one baryon--anti-baryon pair during a single fragmentation. Therefore, this may suggest the need to modify particle production mechanisms and/or the modification of fragmentation functions in Monte Carlo models.
            

\newenvironment{acknowledgement}{\relax}{\relax}
\begin{acknowledgement}
\section*{Acknowledgements}

The ALICE Collaboration would like to thank all its engineers and technicians for their invaluable contributions to the construction of the experiment and the CERN accelerator teams for the outstanding performance of the LHC complex.
The ALICE Collaboration gratefully acknowledges the resources and support provided by all Grid centres and the Worldwide LHC Computing Grid (WLCG) collaboration.
The ALICE Collaboration acknowledges the following funding agencies for their support in building and running the ALICE detector:
A. I. Alikhanyan National Science Laboratory (Yerevan Physics Institute) Foundation (ANSL), State Committee of Science and World Federation of Scientists (WFS), Armenia;
Austrian Academy of Sciences and Nationalstiftung f\"{u}r Forschung, Technologie und Entwicklung, Austria;
Conselho Nacional de Desenvolvimento Cient\'{\i}fico e Tecnol\'{o}gico (CNPq), Universidade Federal do Rio Grande do Sul (UFRGS), Financiadora de Estudos e Projetos (Finep) and Funda\c{c}\~{a}o de Amparo \`{a} Pesquisa do Estado de S\~{a}o Paulo (FAPESP), Brazil;
Ministry of Science \& Technology of China (MSTC), National Natural Science Foundation of China (NSFC) and Ministry of Education of China (MOEC) , China;
Ministry of Science, Education and Sport and Croatian Science Foundation, Croatia;
Ministry of Education, Youth and Sports of the Czech Republic, Czech Republic;
The Danish Council for Independent Research | Natural Sciences, the Carlsberg Foundation and Danish National Research Foundation (DNRF), Denmark;
Helsinki Institute of Physics (HIP), Finland;
Commissariat \`{a} l'Energie Atomique (CEA) and Institut National de Physique Nucl\'{e}aire et de Physique des Particules (IN2P3) and Centre National de la Recherche Scientifique (CNRS), France;
Bundesministerium f\"{u}r Bildung, Wissenschaft, Forschung und Technologie (BMBF) and GSI Helmholtzzentrum f\"{u}r Schwerionenforschung GmbH, Germany;
Ministry of Education, Research and Religious Affairs, Greece;
National Research, Development and Innovation Office, Hungary;
Department of Atomic Energy Government of India (DAE) and Council of Scientific and Industrial Research (CSIR), New Delhi, India;
Indonesian Institute of Science, Indonesia;
Centro Fermi - Museo Storico della Fisica e Centro Studi e Ricerche Enrico Fermi and Istituto Nazionale di Fisica Nucleare (INFN), Italy;
Institute for Innovative Science and Technology , Nagasaki Institute of Applied Science (IIST), Japan Society for the Promotion of Science (JSPS) KAKENHI and Japanese Ministry of Education, Culture, Sports, Science and Technology (MEXT), Japan;
Consejo Nacional de Ciencia (CONACYT) y Tecnolog\'{i}a, through Fondo de Cooperaci\'{o}n Internacional en Ciencia y Tecnolog\'{i}a (FONCICYT) and Direcci\'{o}n General de Asuntos del Personal Academico (DGAPA), Mexico;
Nationaal instituut voor subatomaire fysica (Nikhef), Netherlands;
The Research Council of Norway, Norway;
Commission on Science and Technology for Sustainable Development in the South (COMSATS), Pakistan;
Pontificia Universidad Cat\'{o}lica del Per\'{u}, Peru;
Ministry of Science and Higher Education and National Science Centre, Poland;
Korea Institute of Science and Technology Information and National Research Foundation of Korea (NRF), Republic of Korea;
Ministry of Education and Scientific Research, Institute of Atomic Physics and Romanian National Agency for Science, Technology and Innovation, Romania;
Joint Institute for Nuclear Research (JINR), Ministry of Education and Science of the Russian Federation and National Research Centre Kurchatov Institute, Russia;
Ministry of Education, Science, Research and Sport of the Slovak Republic, Slovakia;
National Research Foundation of South Africa, South Africa;
Centro de Aplicaciones Tecnol\'{o}gicas y Desarrollo Nuclear (CEADEN), Cubaenerg\'{\i}a, Cuba, Ministerio de Ciencia e Innovacion and Centro de Investigaciones Energ\'{e}ticas, Medioambientales y Tecnol\'{o}gicas (CIEMAT), Spain;
Swedish Research Council (VR) and Knut \& Alice Wallenberg Foundation (KAW), Sweden;
European Organization for Nuclear Research, Switzerland;
National Science and Technology Development Agency (NSDTA), Suranaree University of Technology (SUT) and Office of the Higher Education Commission under NRU project of Thailand, Thailand;
Turkish Atomic Energy Agency (TAEK), Turkey;
National Academy of  Sciences of Ukraine, Ukraine;
Science and Technology Facilities Council (STFC), United Kingdom;
National Science Foundation of the United States of America (NSF) and United States Department of Energy, Office of Nuclear Physics (DOE NP), United States of America.    
\end{acknowledgement}

\bibliographystyle{utphys}   
\bibliography{bibliography}

\newpage
\appendix
\section{The ALICE Collaboration}
\label{app:collab}



\begingroup
\small
\begin{flushleft}
J.~Adam$^\textrm{\scriptsize 39}$,
D.~Adamov\'{a}$^\textrm{\scriptsize 87}$,
M.M.~Aggarwal$^\textrm{\scriptsize 91}$,
G.~Aglieri Rinella$^\textrm{\scriptsize 35}$,
M.~Agnello$^\textrm{\scriptsize 31}$\textsuperscript{,}$^\textrm{\scriptsize 113}$,
N.~Agrawal$^\textrm{\scriptsize 48}$,
Z.~Ahammed$^\textrm{\scriptsize 137}$,
S.~Ahmad$^\textrm{\scriptsize 18}$,
S.U.~Ahn$^\textrm{\scriptsize 70}$,
S.~Aiola$^\textrm{\scriptsize 141}$
% ...
\renewcommand\labelenumi{\textsuperscript{\theenumi}~}

\section*{Affiliation notes}
\renewcommand\theenumi{\roman{enumi}}
\begin{Authlist}
\item \Adef{0}Deceased
\item \Adef{idp1806336}{Also at: Georgia State University, Atlanta, Georgia, United States}
\item \Adef{idp3237568}{Also at: Also at Department of Applied Physics, Aligarh Muslim University, Aligarh, India}
\item \Adef{idp4010000}{Also at: M.V. Lomonosov Moscow State University, D.V. Skobeltsyn Institute of Nuclear, Physics, Moscow, Russia}
\end{Authlist}

\section*{Collaboration Institutes}
\renewcommand\theenumi{\arabic{enumi}~}

$^{1}$A.I. Alikhanyan National Science Laboratory (Yerevan Physics Institute) Foundation, Yerevan, Armenia
\\
$^{2}$Benem\'{e}rita Universidad Aut\'{o}noma de Puebla, Puebla, Mexico
\\
$^{3}$Bogolyubov Institute for Theoretical Physics, Kiev, Ukraine
\\
$^{4}$Bose Institute, Department of Physics 
and Centre for Astroparticle Physics and Space Science (CAPSS), Kolkata, India
\\
$^{5}$Budker Institute for Nuclear Physics, Novosibirsk, Russia
\\
$^{6}$California Polytechnic State University, San Luis Obispo, California, United States
\\
$^{7}$Central China Normal University, Wuhan, China
\\
$^{8}$Centre de Calcul de l'IN2P3, Villeurbanne, Lyon, France
\\
$^{9}$Centro de Aplicaciones Tecnol\'{o}gicas y Desarrollo Nuclear (CEADEN), Havana, Cuba
\\
$^{10}$Centro de Investigaciones Energ\'{e}ticas Medioambientales y Tecnol\'{o}gicas (CIEMAT), Madrid, Spain
\\
$^{11}$Centro de Investigaci\'{o}n y de Estudios Avanzados (CINVESTAV), Mexico City and M\'{e}rida, Mexico
\\
$^{12}$Centro Fermi - Museo Storico della Fisica e Centro Studi e Ricerche ``Enrico Fermi', Rome, Italy
\\
$^{13}$Chicago State University, Chicago, Illinois, United States
\\
$^{14}$China Institute of Atomic Energy, Beijing, China
\\
$^{15}$Commissariat \`{a} l'Energie Atomique, IRFU, Saclay, France
\\
$^{16}$COMSATS Institute of Information Technology (CIIT), Islamabad, Pakistan
\\
$^{17}$Departamento de F\'{\i}sica de Part\'{\i}culas and IGFAE, Universidad de Santiago de Compostela, Santiago de Compostela, Spain
\\
$^{18}$Department of Physics, Aligarh Muslim University, Aligarh, India
\\
$^{19}$Department of Physics, Ohio State University, Columbus, Ohio, United States
\\
$^{20}$Department of Physics, Sejong University, Seoul, South Korea
\\
$^{21}$Department of Physics, University of Oslo, Oslo, Norway
\\
$^{22}$Department of Physics and Technology, University of Bergen, Bergen, Norway
\\
$^{23}$Dipartimento di Fisica dell'Universit\`{a} 'La Sapienza'
and Sezione INFN, Rome, Italy
\\
$^{24}$Dipartimento di Fisica dell'Universit\`{a}
and Sezione INFN, Cagliari, Italy
\\
$^{25}$Dipartimento di Fisica dell'Universit\`{a}
and Sezione INFN, Trieste, Italy
\\
$^{26}$Dipartimento di Fisica dell'Universit\`{a}
and Sezione INFN, Turin, Italy
\\
$^{27}$Dipartimento di Fisica e Astronomia dell'Universit\`{a}
and Sezione INFN, Bologna, Italy
\\
$^{28}$Dipartimento di Fisica e Astronomia dell'Universit\`{a}
and Sezione INFN, Catania, Italy
\\
$^{29}$Dipartimento di Fisica e Astronomia dell'Universit\`{a}
and Sezione INFN, Padova, Italy
\\
$^{30}$Dipartimento di Fisica `E.R.~Caianiello' dell'Universit\`{a}
and Gruppo Collegato INFN, Salerno, Italy
\\
$^{31}$Dipartimento DISAT del Politecnico and Sezione INFN, Turin, Italy
\\
$^{32}$Dipartimento di Scienze e Innovazione Tecnologica dell'Universit\`{a} del Piemonte Orientale and INFN Sezione di Torino, Alessandria, Italy
\\
$^{33}$Dipartimento Interateneo di Fisica `M.~Merlin'
and Sezione INFN, Bari, Italy
\\
$^{34}$Division of Experimental High Energy Physics, University of Lund, Lund, Sweden
\\
$^{35}$European Organization for Nuclear Research (CERN), Geneva, Switzerland
\\
$^{36}$Excellence Cluster Universe, Technische Universit\"{a}t M\"{u}nchen, Munich, Germany
\\
$^{37}$Faculty of Engineering, Bergen University College, Bergen, Norway
\\
$^{38}$Faculty of Mathematics, Physics and Informatics, Comenius University, Bratislava, Slovakia
\\
$^{39}$Faculty of Nuclear Sciences and Physical Engineering, Czech Technical University in Prague, Prague, Czech Republic
\\
$^{40}$Faculty of Science, P.J.~\v{S}af\'{a}rik University, Ko\v{s}ice, Slovakia
\\
$^{41}$Faculty of Technology, Buskerud and Vestfold University College, Tonsberg, Norway
\\
$^{42}$Frankfurt Institute for Advanced Studies, Johann Wolfgang Goethe-Universit\"{a}t Frankfurt, Frankfurt, Germany
\\
$^{43}$Gangneung-Wonju National University, Gangneung, South Korea
\\
$^{44}$Gauhati University, Department of Physics, Guwahati, India
\\
$^{45}$Helmholtz-Institut f\"{u}r Strahlen- und Kernphysik, Rheinische Friedrich-Wilhelms-Universit\"{a}t Bonn, Bonn, Germany
\\
$^{46}$Helsinki Institute of Physics (HIP), Helsinki, Finland
\\
$^{47}$Hiroshima University, Hiroshima, Japan
\\
$^{48}$Indian Institute of Technology Bombay (IIT), Mumbai, India
\\
$^{49}$Indian Institute of Technology Indore, Indore, India
\\
$^{50}$Indonesian Institute of Sciences, Jakarta, Indonesia
\\
$^{51}$Inha University, Incheon, South Korea
\\
$^{52}$Institut de Physique Nucl\'eaire d'Orsay (IPNO), Universit\'e Paris-Sud, CNRS-IN2P3, Orsay, France
\\
$^{53}$Institute for Nuclear Research, Academy of Sciences, Moscow, Russia
\\
$^{54}$Institute for Subatomic Physics of Utrecht University, Utrecht, Netherlands
\\
$^{55}$Institute for Theoretical and Experimental Physics, Moscow, Russia
\\
$^{56}$Institute of Experimental Physics, Slovak Academy of Sciences, Ko\v{s}ice, Slovakia
\\
$^{57}$Institute of Physics, Academy of Sciences of the Czech Republic, Prague, Czech Republic
\\
$^{58}$Institute of Physics, Bhubaneswar, India
\\
$^{59}$Institute of Space Science (ISS), Bucharest, Romania
\\
$^{60}$Institut f\"{u}r Informatik, Johann Wolfgang Goethe-Universit\"{a}t Frankfurt, Frankfurt, Germany
\\
$^{61}$Institut f\"{u}r Kernphysik, Johann Wolfgang Goethe-Universit\"{a}t Frankfurt, Frankfurt, Germany
\\
$^{62}$Institut f\"{u}r Kernphysik, Westf\"{a}lische Wilhelms-Universit\"{a}t M\"{u}nster, M\"{u}nster, Germany
\\
$^{63}$Instituto de Ciencias Nucleares, Universidad Nacional Aut\'{o}noma de M\'{e}xico, Mexico City, Mexico
\\
$^{64}$Instituto de F\'{i}sica, Universidade Federal do Rio Grande do Sul (UFRGS), Porto Alegre, Brazil
\\
$^{65}$Instituto de F\'{\i}sica, Universidad Nacional Aut\'{o}noma de M\'{e}xico, Mexico City, Mexico
\\
$^{66}$Institut Pluridisciplinaire Hubert Curien (IPHC), Universit\'{e} de Strasbourg, CNRS-IN2P3, Strasbourg, France
\\
$^{67}$iThemba LABS, National Research Foundation, Somerset West, South Africa
\\
$^{68}$Joint Institute for Nuclear Research (JINR), Dubna, Russia
\\
$^{69}$Konkuk University, Seoul, South Korea
\\
$^{70}$Korea Institute of Science and Technology Information, Daejeon, South Korea
\\
$^{71}$KTO Karatay University, Konya, Turkey
\\
$^{72}$Laboratoire de Physique Corpusculaire (LPC), Clermont Universit\'{e}, Universit\'{e} Blaise Pascal, CNRS--IN2P3, Clermont-Ferrand, France
\\
$^{73}$Laboratoire de Physique Subatomique et de Cosmologie, Universit\'{e} Grenoble-Alpes, CNRS-IN2P3, Grenoble, France
\\
$^{74}$Laboratori Nazionali di Frascati, INFN, Frascati, Italy
\\
$^{75}$Laboratori Nazionali di Legnaro, INFN, Legnaro, Italy
\\
$^{76}$Lawrence Berkeley National Laboratory, Berkeley, California, United States
\\
$^{77}$Moscow Engineering Physics Institute, Moscow, Russia
\\
$^{78}$Nagasaki Institute of Applied Science, Nagasaki, Japan
\\
$^{79}$National and Kapodistrian University of Athens, Physics Department, Athens, Greece, Athens, Greece
\\
$^{80}$National Centre for Nuclear Studies, Warsaw, Poland
\\
$^{81}$National Institute for Physics and Nuclear Engineering, Bucharest, Romania
\\
$^{82}$National Institute of Science Education and Research, Bhubaneswar, India
\\
$^{83}$National Research Centre Kurchatov Institute, Moscow, Russia
\\
$^{84}$Niels Bohr Institute, University of Copenhagen, Copenhagen, Denmark
\\
$^{85}$Nikhef, Nationaal instituut voor subatomaire fysica, Amsterdam, Netherlands
\\
$^{86}$Nuclear Physics Group, STFC Daresbury Laboratory, Daresbury, United Kingdom
\\
$^{87}$Nuclear Physics Institute, Academy of Sciences of the Czech Republic, \v{R}e\v{z} u Prahy, Czech Republic
\\
$^{88}$Oak Ridge National Laboratory, Oak Ridge, Tennessee, United States
\\
$^{89}$Petersburg Nuclear Physics Institute, Gatchina, Russia
\\
$^{90}$Physics Department, Creighton University, Omaha, Nebraska, United States
\\
$^{91}$Physics Department, Panjab University, Chandigarh, India
\\
$^{92}$Physics Department, University of Cape Town, Cape Town, South Africa
\\
$^{93}$Physics Department, University of Jammu, Jammu, India
\\
$^{94}$Physics Department, University of Rajasthan, Jaipur, India
\\
$^{95}$Physikalisches Institut, Eberhard Karls Universit\"{a}t T\"{u}bingen, T\"{u}bingen, Germany
\\
$^{96}$Physikalisches Institut, Ruprecht-Karls-Universit\"{a}t Heidelberg, Heidelberg, Germany
\\
$^{97}$Physik Department, Technische Universit\"{a}t M\"{u}nchen, Munich, Germany
\\
$^{98}$Purdue University, West Lafayette, Indiana, United States
\\
$^{99}$Pusan National University, Pusan, South Korea
\\
$^{100}$Research Division and ExtreMe Matter Institute EMMI, GSI Helmholtzzentrum f\"ur Schwerionenforschung GmbH, Darmstadt, Germany
\\
$^{101}$Rudjer Bo\v{s}kovi\'{c} Institute, Zagreb, Croatia
\\
$^{102}$Russian Federal Nuclear Center (VNIIEF), Sarov, Russia
\\
$^{103}$Saha Institute of Nuclear Physics, Kolkata, India
\\
$^{104}$School of Physics and Astronomy, University of Birmingham, Birmingham, United Kingdom
\\
$^{105}$Secci\'{o}n F\'{\i}sica, Departamento de Ciencias, Pontificia Universidad Cat\'{o}lica del Per\'{u}, Lima, Peru
\\
$^{106}$Sezione INFN, Bari, Italy
\\
$^{107}$Sezione INFN, Bologna, Italy
\\
$^{108}$Sezione INFN, Cagliari, Italy
\\
$^{109}$Sezione INFN, Catania, Italy
\\
$^{110}$Sezione INFN, Padova, Italy
\\
$^{111}$Sezione INFN, Rome, Italy
\\
$^{112}$Sezione INFN, Trieste, Italy
\\
$^{113}$Sezione INFN, Turin, Italy
\\
$^{114}$SSC IHEP of NRC Kurchatov institute, Protvino, Russia
\\
$^{115}$Stefan Meyer Institut f\"{u}r Subatomare Physik (SMI), Vienna, Austria
\\
$^{116}$SUBATECH, Ecole des Mines de Nantes, Universit\'{e} de Nantes, CNRS-IN2P3, Nantes, France
\\
$^{117}$Suranaree University of Technology, Nakhon Ratchasima, Thailand
\\
$^{118}$Technical University of Ko\v{s}ice, Ko\v{s}ice, Slovakia
\\
$^{119}$Technical University of Split FESB, Split, Croatia
\\
$^{120}$The Henryk Niewodniczanski Institute of Nuclear Physics, Polish Academy of Sciences, Cracow, Poland
\\
$^{121}$The University of Texas at Austin, Physics Department, Austin, Texas, United States
\\
$^{122}$Universidad Aut\'{o}noma de Sinaloa, Culiac\'{a}n, Mexico
\\
$^{123}$Universidade de S\~{a}o Paulo (USP), S\~{a}o Paulo, Brazil
\\
$^{124}$Universidade Estadual de Campinas (UNICAMP), Campinas, Brazil
\\
$^{125}$Universidade Federal do ABC, Santo Andre, Brazil
\\
$^{126}$University of Houston, Houston, Texas, United States
\\
$^{127}$University of Jyv\"{a}skyl\"{a}, Jyv\"{a}skyl\"{a}, Finland
\\
$^{128}$University of Liverpool, Liverpool, United Kingdom
\\
$^{129}$University of Tennessee, Knoxville, Tennessee, United States
\\
$^{130}$University of the Witwatersrand, Johannesburg, South Africa
\\
$^{131}$University of Tokyo, Tokyo, Japan
\\
$^{132}$University of Tsukuba, Tsukuba, Japan
\\
$^{133}$University of Zagreb, Zagreb, Croatia
\\
$^{134}$Universit\'{e} de Lyon, Universit\'{e} Lyon 1, CNRS/IN2P3, IPN-Lyon, Villeurbanne, Lyon, France
\\
$^{135}$Universit\`{a} di Brescia, Brescia, Italy
\\
$^{136}$V.~Fock Institute for Physics, St. Petersburg State University, St. Petersburg, Russia
\\
$^{137}$Variable Energy Cyclotron Centre, Kolkata, India
\\
$^{138}$Warsaw University of Technology, Warsaw, Poland
\\
$^{139}$Wayne State University, Detroit, Michigan, United States
\\
$^{140}$Wigner Research Centre for Physics, Hungarian Academy of Sciences, Budapest, Hungary
\\
$^{141}$Yale University, New Haven, Connecticut, United States
\\
$^{142}$Yonsei University, Seoul, South Korea
\\
$^{143}$Zentrum f\"{u}r Technologietransfer und Telekommunikation (ZTT), Fachhochschule Worms, Worms, Germany
\endgroup

\end{document}